%% file: main.tex
 \documentclass[final,3p,times]{elsarticle}

\usepackage{lipsum} 

\usepackage{amssymb}
\usepackage{amsfonts}
\usepackage{amsmath}
\usepackage{amssymb}
\usepackage{bm,amsbsy} 
\usepackage{algorithm,algorithmic}

 \usepackage{lineno}
\usepackage{subcaption}
\usepackage{hyperref}

\newcommand{\Nrm}{\mathcal{N}}
\newcommand{\Tr}{\textrm{Tr}}

\newcommand{\vx}{\mathbf{x}}

\newcommand{\vw}{\mathbf{w}}

\newcommand{\vy}{\mathbf{y}}
\newcommand{\trp}{{^T}} 

\newcommand{\inv}{^{-1}}

\newcommand{\vtheta}{\mathbf{\ensuremath{\bm{\theta}}}}

\newcommand{\vu}{\mathbf{u}}

\newcommand{\vones}{\mathbf{1}}

\newcommand{\nsevar}{\sigma^2}

\newcommand{\RR}{\mathbb{R}}

\DeclareMathOperator{\argmin}{argmin}

\newcommand{\vecop}{\mathrm{vec}}
\newcommand{\real}[1]{\ensuremath{\mathbb{R}^{{#1}}}}

\newcommand{\mat}[1]{\ensuremath{\mathbf{#1}}}
\newcommand{\mats}[2]{\ensuremath{\mathbf{#1}_{#2}}}

\newcommand{\ves}[2]{\ensuremath{\mathbf{#1}_{#2}}}

\newcommand{\vehs}[2]{\ensuremath{\mathbf{\hat #1}_{#2}}}

\newcommand{\normaldist}[2]{\ensuremath{\mathcal{N}\left(#1,#2\right)}}

\newcommand{\normtwo}[1]{\ensuremath{\|{#1}\|_2}}
\newcommand{\normfro}[1]{\ensuremath{\|{#1}\|_F}}

\newcommand\blfootnote[1]{%
  \begingroup
  \renewcommand\thefootnote{}\footnote{#1}%
  \addtocounter{footnote}{-1}%
  \endgroup
}

\journal{Neuropsychologia}

\begin{document}

\begin{frontmatter}



\title{Incorporating structured assumptions with probabilistic graphical models in fMRI data analysis}


\author[label1,label2]{Ming Bo Cai\corref{cor1}\thanks{contrib}}
\author[label3]{Michael Shvartsman\thanks{contrib}}
\author[label4]{Anqi Wu\thanks{contrib}}
\author[label5]{Hejia Zhang\thanks{contrib}}
\author[label6]{Xia Zhu\thanks{contrib}}
\address[label1]{International Research Center for Neurointelligence (WPI-IRCN), UTIAS, The University of Tokyo, Japan}
\address[label2]{Princeton Neuroscience Institute, Princeton University, United States}
\address[label3]{Facebook Reality Labs, United States}
\address[label4]{Mortimer B. Zuckerman Mind Brain Behavior Institute, Columbia University, United States}
\address[label5]{Department of Electrical Engineering, Princeton University, United States}
\address[label6]{Intel Corporation, United States}
\cortext[cor1]{Corresponding author: mingbo.cai@ircn.jp}
\cortext[contrib]{All authors contributed equally.}

\begin{abstract}
\blfootnote{\textsuperscript{\textcopyright} 2020. This manuscript version is made available under the CC-BY-NC-ND 4.0 license http://creativecommons.org/licenses/by-nc-nd/4.0/}
With the wide adoption of functional magnetic resonance imaging (fMRI) by cognitive neuroscience researchers, large volumes of brain imaging data have been accumulated in recent years. Aggregating these data to derive scientific insights often faces the challenge that fMRI data are high-dimensional, heterogeneous across people, and noisy. These challenges demand the development of computational tools that are tailored both for the neuroscience questions and for the properties of the data. We review a few recently developed algorithms in various domains of fMRI research: fMRI in naturalistic tasks, analyzing full-brain functional connectivity, pattern classification, inferring representational similarity and modeling structured residuals. These algorithms all tackle the challenges in fMRI similarly: they start by making clear statements of assumptions about neural data and existing domain knowledge, incorporating those assumptions and domain knowledge into probabilistic graphical models, and using those models to estimate properties of interest or latent structures in the data. Such approaches can avoid erroneous findings, reduce the impact of noise, better utilize known properties of the data, and better aggregate data across groups of subjects. With these successful cases, we advocate wider adoption of explicit model construction in cognitive neuroscience. Although we focus on fMRI, the principle illustrated here is generally applicable to brain data of other modalities.

\end{abstract}


\end{frontmatter}


\section{Introduction}
\label{S:1}
Functional magnetic resonance imaging (fMRI) \cite{ogawa1990brain,belliveau1991functional} is a powerful tool to study the brain's activity and functions. The fluctuation of the fMRI signal is related to the fluctuation of the concentrations of the oxygenated and deoxygenated hemoglobin in the blood, which follows the increase or decrease of local neuronal activity with a delay \cite{buxton2013physics,heeger2002does}. This relation to the neural activity, together with its non-invasive nature, full brain coverage and reasonable balance between spatial and temporal resolution, makes fMRI a widely used brain imaging technique for studying the neural correlates of perceptual and cognitive processes in humans. 

\begin{figure}[!ht]
  \includegraphics[width=\linewidth]{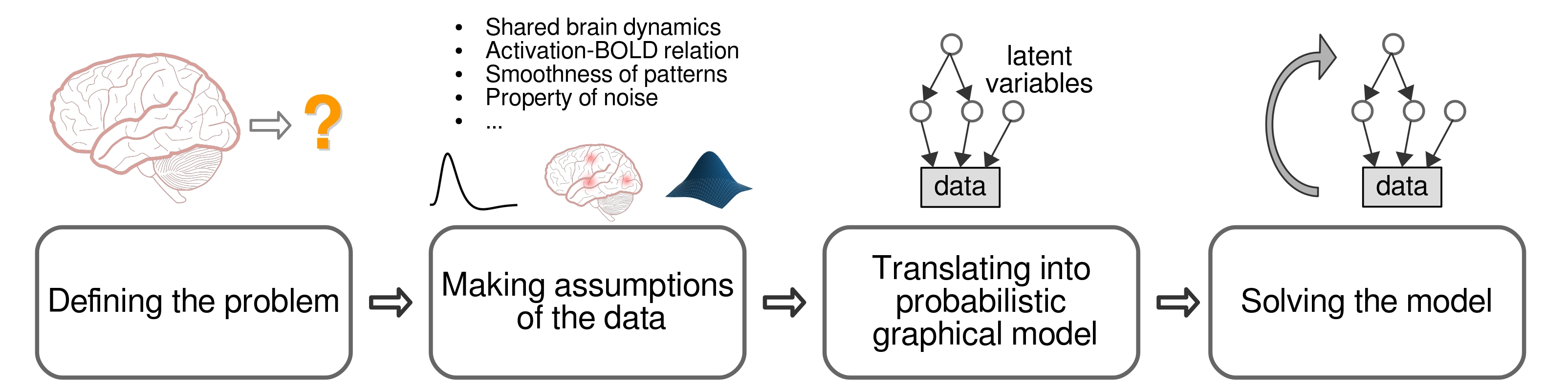}
  \captionof{figure}{The PGM-based approach to analyze neural data. In general, this involves four steps: (1) clearly defining the problem to solve or the question being asked; (2) making assumptions about the property of the data, including domain knowledge about data and causal relation between latent variables and measured data; (3) translating these assumptions to a probabilistic graphical model which expresses how latent variables together generate measured data. The model uses conditional probability distributions between variables to capture their causal relations; (4) solving the model to infer latent variables or to draw conclusion for the question being asked in the first step.}
  
  \label{fig:modeling_approach}
\end{figure}

However, deriving insights about neural information processing from fMRI data can be challenging in many situations, because (1) fMRI signals only indirectly relate to neural activity \cite{buxton2013physics,friston1995characterizing}; (2) the data typically contain noise and unknown physiological signals with complex spatial and temporal correlation, and various artifacts \cite{triantafyllou2005comparison,bright2015fmri,ZarahnAguirreDEsposito1997}; (3) the number of brain volumes scanned in each experiment is much smaller than the number of voxels (high dimensionality of data in contrast to small sample size); and (4) there is large variation in detailed brain anatomical structures and functional organization across people \cite{finn2016individual,suarez2020linking}, making it harder to aggregate data across people. Analysis tools need to take these factors into account in order to obtain fruitful insight from data. One promising approach to address these challenges is that of probabilistic graphical models. Probabilistic graphical models (PGM) are frameworks used to create probabilistic models of complex data distributions and represent them in compact graphical representation \cite{koller2009probabilistic}. PGMs have been widely used in many different fields ~\cite{friedman2004inferring,ahelegbey2016econometrics,ji2019probabilistic}. In behavioral studies of perception and cognition, PGM is not only a framework for describing the computational processes in the brain \cite{ma2012organizing,l2008bayesian,geisler2011contributions}, but also a framework for testing different models against behavior \cite{kruschke2010believe,shiffrin2008survey,etz2018introduction}. However, except in few domains, its value for neural imaging analysis has not been fully appreciated. With PGMs, we can explicitly assume the relations and dependencies between quantities of interest and the data, construct hierarchical generative models that posit how fMRI data is generated from mental processes, incorporate structural assumptions about the data and domain knowledge into the models, and take into account the uncertainty/noise in fMRI data explicitly. These are benefits that are hard to achieve in conventional fMRI analysis tools, which provide a universal set of analyses whose baseline assumptions may or may not apply to specific datasets. Fig.~\ref{fig:modeling_approach} describes the scheme of building PGMs for fMRI study. In this paper, we illustrate how to build PGMs to solve neural imaging analysis problems following this scheme.

As illustrated in Fig.~\ref{fig:modeling_approach}, an approach which relies on explicit PGM typically involves four major steps. The first step is defining the problem: deciding what question is asked or what problem needs to be solved, and deciding what quantity allows one to answer the question or to characterize certain aspect of the brain. This is essentially the hypothesis generation step of hypothesis-driven science, but we emphasize it as a distinct step because the hypothesis needs to be precise enough to translate into a model in subsequenct steps. 

After the question is clearly defined, the second step is to make explicit assumptions of how the quantity of interest and experimental manipulations directly or indirectly contribute to the data to be analyzed, and to make assumptions of how variables of no interest (nuisance factors) may jointly impact the data. If there is domain knowledge of the properties of fMRI data that can help construct models of the data, it should be clearly stated at this step as well. 

The third step is to translate these assumptions and domain knowledge into a computational model. Such computational models can often be described by probabilistic graphical models (PGM) \cite{koller2009probabilistic} composed of nodes and directed edges between nodes. The nodes and edges together form a graph. When building PGMs, the data, experimental manipulation, quantity of interest and nuisance factors that are considered in the assumptions of the previous step all become variables (either known or unknown) and are each represented by a node in the graph. The hypothesized relations between variables in the models are expressed as conditional probability of one variable given one or more other variables, and are represented by directed edges. Each edge is directed from one variable to another variable that is conditioned on it (i.e., the distribution of the variable at the head of an edge (arrow) depends on the variable at the tail of the edge). The domain knowledge is either captured by the prior distribution of certain variables in the graph, or in the form of the conditional dependencies. The probabilistic nature of such models makes them a natural choice for capturing the noise properties in the system and the potential uncertainty in the estimates of parameters from the researchers' perspective. 

Once the PGM is built, the fourth step is to deploy computational techniques to estimate the unknown variables of interest in the model. This step essentially inverts the model by inferring variables of interest at the source of the directed edges in the graphical model. In some cases, inferring these variables serves to answer the original question by providing characterization of some aspect of brain activity. In other cases, when the scientific question is to test competing hypotheses, the competing hypotheses should be translated into PGMs that differ in either the range of values of some key variables or in the structures of the models. The selection of the winning model can be either based on classical statistical tests of the inferred values of the key variables (when each value is inferred independently), or based on the likelihood that each model can give rise to the data (model evidence), marginalizing unknown variables \cite{mackay2003information,jeffreys1998theory}. To approximate posterior distributions of latent variables (variables that are directly or indirectly causal to the nodes representing observable data) in the probabilistic graphical models given the observed data, techniques such as Markov Chain Monte Carlo (MCMC) \cite{metropolis1953equation,hastings1970monte} or variational Bayes \cite{jordan1999introduction} are often employed \cite{gelman2013bayesian}. In certain cases, when the posterior distribution of these latent variables can be analytically derived, exact inference of the posterior distribution or the \textit{maximum a posteriori} values of the variables can often be achieved. A full discussion of the inference methods is out of the scope of this paper. Interested readers may refer to tutorials such as Chapter 8 of \cite{bishop2006pattern} or part II of \cite{koller2009probabilistic}.

Because a PGM is explicitly built, it is easy to evaluate whether the inference procedure can reliably recover the variables of interest in the model, by simulating data according to the model and comparing the recovered values of those variables with the values used in the simulation. In contrast, traditional approaches 
without building an explicit model of the data generating process lack the ability to simulate data in accordance with its (implicit) assumptions. Without simulating data, it becomes impossible to verify that an analysis can yield correct results, because researchers are only left with real neural data of which the generative process and the ground truth of the variables of interest are not known. Thus, there is no guarantee that the quantity extracted by analysis methods without explicit assumptions of data generating process bears direct relation to what the researchers are interested in.

In addition to transparency and verifiability, PGMs
offer the flexibility to combine the advantage of various pieces of domain knowledge of the brain (for example, brain activation patterns tend to be spatially smooth). This is because domain knowledge can be translated into a prior distribution of certain form over some latent variables in the PGM. With the PGM as a backbone, different prior distributions may act as add-on parts that can be plugged in at different places of the model, depending on what domain knowledge is proper for the purpose of analysis. For example, in \ref{S:ivy}, we show that the smoothness assumption of the brain activity and the similarity of brain networks across people are incorporated as the 3D Gaussian shape of the spatial basis for brain patterns and the Gaussian distribution of node location across subjects, respectively. In \ref{S:anqi}, we show that two types of prior knowledge about fMRI decoding weights, smoothness and sparsity, can also be incorporated together by assuming a Gaussian process prior on the joint distribution of the fMRI decoding weights of all voxels. 


In the following, we select example analysis methods developed in different research domains to illustrate how the PGM approach to neural imaging data can be applied to discovering shared neural dynamics across participants doing the same task, modeling the functional connectivity among brain regions, improving the performance of decoding mental contents and obtaining more biologically informed decoding weights, reducing the bias in estimating similarity among activation patterns, and providing more comprehensive model of the noise in fMRI data. 
These methods together illustrate how the PGM \cite{koller2009probabilistic} can accommodate domain knowledge and known properties of the data and facilitate aggregating information over larger datasets. These features allow us to mitigate the limitations in fMRI data: high dimensionality (many voxels), low sample size in single participant, heterogeneity across participants and high noise. Although introducing a PGM is not the only way to overcome these limitations, it is one of the simplest ways to achieve transparency, verifiability and interpretability, due to explicit modeling.

Because the focus is on illustrating the principle of PGMs, this paper can by no means provide a thorough review of all PGM-based methods of fMRI analysis. For example, the use of dynamic causal modeling
\cite{friston2003dynamic,stephan2010ten} to infer the interactive relations among brain regions, the construction of encoding models \cite{naselaris2009bayesian,nishimoto2011reconstructing,naselaris2011encoding,dumoulin2008population} to understand the features encoded by a brain region or to reconstruct perceived sensory inputs, and the development of a probabilistic event segmentation model \cite{baldassano2017discovering} to discover distinctive and sustained brain states, are all good illustrations of the four-step procedure above. Readers are encouraged to also refer to several other reviews (e.g., \cite{woolrich2012bayesian,woolrich2009bayesian}) on Bayesian approaches to fMRI for a more comprehensive understanding of other existing PGM tools that share the advantages illustrated here. 

While we advocate for explicit construction of probabilistic models in fMRI analysis, PGM-based methods do still have limitations. For example, performing efficient inference on PGMs can still scale poorly in both computation and memory, limiting their use on large-scale data without specially-tuned algorithms and approximations. Furthermore, deriving the inference algorithm for any specific model has historically required extensive knowledge in computer science or statistics, and it is not always easy to establish how robust any specific model is to mismatches between the assumptions made in the models and the true data properties. But the overall outlook is encouraging: algorithmic improvements have enabled scaling up models previously considered impossible (e.g.\ \cite{Wang2019}), general tools for probabilistic inference have blurred the lines between practitioner and methods developer (e.g.\ \cite{Carpenter2016,Salvatier2016,Bingham2019}), and advances in these aspects of PGM are an area of active research that will continue within and outside the neuroscientific domain.







\section{Examples of PGM-based analysis methods for fMRI data}
\label{S:2}

\subsection{Discovering latent neural dynamics for naturalistic task}
\label{S:hejia}
\input{hejia.tex}



\subsection{Discovering full-brain functional connectivity from fMRI}
\label{S:ivy}
\input{ivy.tex}

\subsection{Obtaining biologically informed decoding weights on fMRI patterns}
\label{S:anqi}

\input{anqi.tex}

\subsection{Inferring representational similarity between neural patterns}
\label{S:mingbo}
\input{mingbo.tex}

\subsection{Modeling structured residuals}
\label{S:mike}
\input{mike.tex}
\section{Discussion}
In this paper, we use five computational tools developed for different goals in fMRI research to illustrate how to build probabilistic graphical models (PGMs) to address important questions arising in neuroimaging studies. 
These methods also illustrate how the PGM \cite{koller2009probabilistic}, which is central to the methods reviewed, can accommodate domain knowledge and known properties of the data and facilitate aggregating information over larger datasets. These features allow us to mitigate the limitations in fMRI data: high dimensionality (many voxels), low sample size in single subject, heterogeneity across subjects and complex noise with high magnitude. The PGM-based approach helps ensure the faithfulness of an algorithm to its original purpose and provides flexibility in model building.

To tackle the limits of high dimensionality and low sample size, SRM \cite{chen2015reduced} uses the existence of a shared latent response as its core assumption, which allows aggregating data from multiple subjects; HTFA~\cite{Htfa18} uses a hierarchical model across subjects to discover common nodes in many brains. By utilizing big data across many subjects, both methods essentially increase the sample size to discover common structure in the data. In addition, the low-rank factor model underlying both methods reduces the model complexity, thus mitigating overfitting.

In aggregating data, both SRM and (H)TFA tolerate the heterogeneity of data across subjects, but in slightly different ways: SRM assumes different spatial weight matrices across subjects while HTFA allows the spatial location of the same node in different subjects to vary. Similarly, an extension of BRSA, the Group BRSA \cite{cai2019representational} allows spatial patterns to differ across subjects while assuming the same similarity matrix is shared by subjects.

An alternative way to mitigate high dimensionality and low sample size is to introduce domain knowledge which trades off between bias and variance in parameter estimation. The three-dimensional Gaussian kernel in (H)TFA~\cite{MannEtal14b,Htfa18} can be considered as adopting the belief that fMRI activations are smooth and local. DRD \citep{wu2019dependent} introduces similar domain knowledge (region sparsity) to tackle the problem by using a Gaussian Process prior on the logarithm of decoding weight variance. This prior allows the weights to have more flexible spatial patterns than Gaussian blobs. Although not reviewed in this article, the method of estimating population receptive field \cite{dumoulin2008population} and more generally, the encoding model approach \cite{naselaris2011encoding} essentially also bring in domain knowledge of neural tuning properties in modeling fMRI data.

Aggregating more data and introducing domain knowledge both essentially reduce the impact of high noise in fMRI data. BRSA and kronecker-separable factor model variants \citep{Shvartsman2018} go one step further by explicitly modeling the spatial and temporal correlation structure in noise. BRSA separates the spatially correlated and independent noise components and models the former with a factor model, allowing  for a more complex correlation structure. The matrix-normal formalism assumes separability of the residual covariance structure into one corresponding to spatial covariance and one corresponding to temporal covariance, largely reducing the number of free parameters while still being able to capture the major structure in noise. Explicitly modeling the noise structure helps reduce bias in estimation arising from the mismatch between an overly simplified noise assumption and the complex property of noise in the data.

In addition to tackling the limitation in fMRI to increase the power for discovering meaningful information in the data, one advantage of the PGM-based approach is its faithfulness to the original goal of a research. This is illustrated in the case of BRSA, where an approach without explicitly examining the data generating and analyzing process may overlook the difference between the output of an early-stage analysis procedure and the true quantity of data that the procedure attempts to estimate, and may introduce spurious results. PGMs allow for simulation of data according to the model and verification of the inference algorithm. This is an advantage not easily achieved by analysis procedures developed without an explicit model. In fact, during sequential applications of analysis or filtering steps, later steps may reintroduce artifacts intended to be removed by early steps \cite{lindquist2019modular}, and variation in complex pipelines may vary the results~\cite{carp2012plurality}. In functional connectivity analysis, various denoising procedures can introduce spurious brain network correlational structures \cite{chen2017nuisance,murphy2017towards,murphy2009impact,saad2012trouble,leonardi2015spurious}. These are often due to the interaction between the preprocessing procedures and later-stage analysis, which is hard to foresee without building and analyzing explicit generative models.

The PGM-based approach to neuroimaging analysis also offers the flexibility of combining advantages of different models and tailoring models for new application domains. This has been illustrated by the extensions of SRM to several variants that utilize partial labels of data \cite{turek2017semi} or datasets with partially overlapping subjects \cite{zhang2018transfer}. Likewise, it is illustrated in the development of separable-covariance variants of existing models \citep{Shvartsman2018}. It is an interesting future research direction to develop new tools that combine the advantage of the existing PGM-based methods, including the models reviewed here. Understanding the commonality among  models is the first step towards integrating them. This is the reason we intentionally use the same notation and matrix orientation of the data matrices in this paper to help readers see the commonality among these methods. Furthermore, several of the tools in this article are available in the same open source package \href{https://brainiak.org/}{\textit{Brain Imaging Analysis Kit} (\textit{BrainIAK})} \cite{kumar2019brainiak}, which makes it easier for tool developers to understand how the computational models and inference algorithms ultimately turn into functioning code and to draw inspiration from these tools (examples for the usage of most algorithms can be found at \href{https://github.com/brainiak/brainiak/tree/master/examples}{https://github.com/brainiak/brainiak/tree/master/examples}). 

Although PGM-based methods come with the aforementioned advantages, they are not without limitations. The first limitation is the speed of computation. Because such models need to consider uncertainty of unknown variables, marginalization of unknown variables is involved, which often requires inverting relatively large matrices, a time- and memory-consuming computation. However, with the advance of parallel computing techniques and code optimization \cite{anderson2016enabling,Gardner2018}, these limitations are gradually being resolved. Second, although the integrative approach of PGM-based methods reduces the chance of obtaining spurious outcomes due to interaction between different stages of data processing, it does reduce the flexibility provided by traditional approach which concatenates many modular analysis tools as a pipeline \cite{esteban2019fmriprep}. Traditional pipeline approaches allow for fast reanalysis when more data (e.g., a new subject) are added, while some of the PGM-based methods may need to redo the analysis on all the data in such situation, or at least require deriving new model update equations. The third limitation is the potential sensitivity of these methods to the correctness of the prior assumptions and generative models used in such methods. For example, noise are often assumed to follow variants of multivariate Gaussian distributions for the easiness of inverting models. But more investigations are needed on the impact of such assumptions when the data distributions in fact violate the assumptions, for example, by having heavier tails than Gaussian distributions. If a model is not a good description of the true generative process of data, then the analysis result may not reflect the ground truth underlying the data. Some assumptions may still be overly simplified compared to the complex nature of true fMRI noise (here, noise may include intrinsic neural activity). This is a challenge but also an opportunity of the PGM-based approach, because making all assumptions explicit makes it easier to examine the impact of the assumptions being made. One may ask how to check whether an assumption of the noise property is correct. Although it is hard to know the true model for data, PGMs at least offer the ability to calculate the likelihood of data given any assumption of noise property. By comparing which assumptions about noise gives higher likelihood for the data, one can decide what assumptions are most appropriate for the data acquired. 
Finally, although PGM-based approaches aim to consider as much of the noise property as possible, they are still not end-to-end, in the sense that various preprocessing procedures, such as slice-timing correction, motion correction, and spatial distortion correction, are still performed separately prior to employing these methods. The generative process of motion-induced artifacts is typically not modeled in these PGM-based methods. Building a PGM which directly models the raw data straight out of fMRI scanners is still too complex a process. One needs to find a balance between the advantage from the explicit and probabilistic nature of PGM and the complexity introduced by modeling every detail of the data.

Beyond the properties of fMRI data that have been discussed in this paper, there are many more complexities in the properties of fMRI data. These complexities may all influence the conclusions one can draw from the data, depending on how much they are taken into account. For example, the temporal profile of haemodynamic response can differ not only across regions, but also across brains. PGM-based methods in fact played an important role in modeling and quantifying this variation \cite{woolrich2004fully}. Future work that incorporates this variation in methods such as BRSA may improve the power of the algorithm and help users evaluate how sensitive their analyses are to such variations. Behavioral contingencies such as reaction time can also influence the fMRI response. Currently their impact is usually modeled deterministically when building design matrices in traditional regression analysis and in BRSA \cite{cai2019representational}. Future work may incorporate the probabilistic impact of such behavioral contingency on fMRI responses, or treat these behavioral measures as additional observations that can be probabilistically predicted by a PGM.
The signal-to-noise ratio of data can be influenced by the speed of fMRI acquisition, together with other factors. So there is a trade-off between data quality and the spatial and temporal resolution of the data. Therefore, there is a need for developing new PGM which simultaneously incorporates multiple types of noise and artifacts, and estimate their relative magnitudes due to the choice of data acquisition protocol. Such work holds promise to improve over existing methods~\cite{ellis2020facilitating} that estimate different noise separately for evaluating the power of an fMRI study. However, as stated above, there is a tradeoff between the amount of details a model can capture and the difficulty of making inference based on a complex model.

In the new era of big data for neuroscience \cite{landhuis2017neuroscience,kandel2013neuroscience,van2014human}, facilitating data sharing is obviously one of the most important effort for making big data analysis possible \cite{poldrack2014making,choudhury2014big,gorgolewski2017openneuro,gorgolewski2016brain}. One step further, developing computational models that derive insights from big data is another key for the field of neuroscience to benefit from increasing data size, which should also be in synergy with developing theories of the essence of the neural computation \cite{cohen2017computational,sejnowski2014putting,bzdok2017inference}. We suggest that future method development places model building at the center of its focus.

\section{Acknowledgement} MBC is supported by National Institute of Drug Abuse award R01DA042065, USA and World Premier International Research Center Initiative (WPI), MEXT, Japan.

\bibliographystyle{model1-num-names}
\bibliography{mingbo,mike,ivy,anqi,hejia}







\end{document}

%% file: hejia.tex
\vspace{0.1in}
\noindent\textsl{- Defining the problem: aggregating multi-subject fMRI data}\\
fMRI datasets with naturalistic stimuli, such as movies or audiobooks, usually have limited number of samples per subject. In general, fMRI datasets not only have a large number of voxels, but also tend to have a small number of time points due to the limitation of samples per experiment session as a result of the slowness of the haemodynamic response and limited sample rate of the scanner. In fMRI datasets with naturalistic stimuli, it is also infeasible to collect many samples from a single subject when the experiments require the natural stimulus to be fresh to the subjects, so each subject could only be exposed to the same stimulus once. 
Therefore, to improve analysis sensitivity, we need to aggregate data from multiple subjects with the same stimulus effectively. The idea is similar to repeated-measures designs in neuroscience where the same variable is measured multiple times, but here the repetition is over different subjects. In our fMRI analysis application, we want to find what is common across subjects. The challenge is that the anatomical and functional structures between subjects are not aligned \cite{talairach1988co}. For example, when listening to the same music, a musician and a person without any music training will probably have different responses. Some early attempts applied pipelines such as averaging the fMRI data from all subjects after anatomical alignment, which assumes voxels of different brains have one-to-one correspondence \cite{talairach1988co, mazziotta2001probabilistic}. In contrast, the Shared Response Model (SRM) \cite{chen2015reduced} is a Bayesian factor analysis model that finds the shared latent neural dynamics across subjects in a multi-subject fMRI dataset after anatomical alignment, without assuming one-to-one voxel correspondence.

\vspace{0.1in}
\noindent\textsl{- Making assumptions: temporally-aligned stimulus}\\
SRM assumes that the stimulus in a naturalistic task dataset is temporally-aligned. That is, all the subjects receive the same stimulus at the same time point in the task. Therefore, we assume that all the subjects share the same low-dimensional latent representation within a dataset, called "shared response." 
On the other hand, to account for the differences between subjects, SRM assumes that each subject has a subject-specific spatial basis for generating the observed fMRI data from the shared response. 

\begin{figure}[!ht]
\centering
\includegraphics[width=\linewidth]{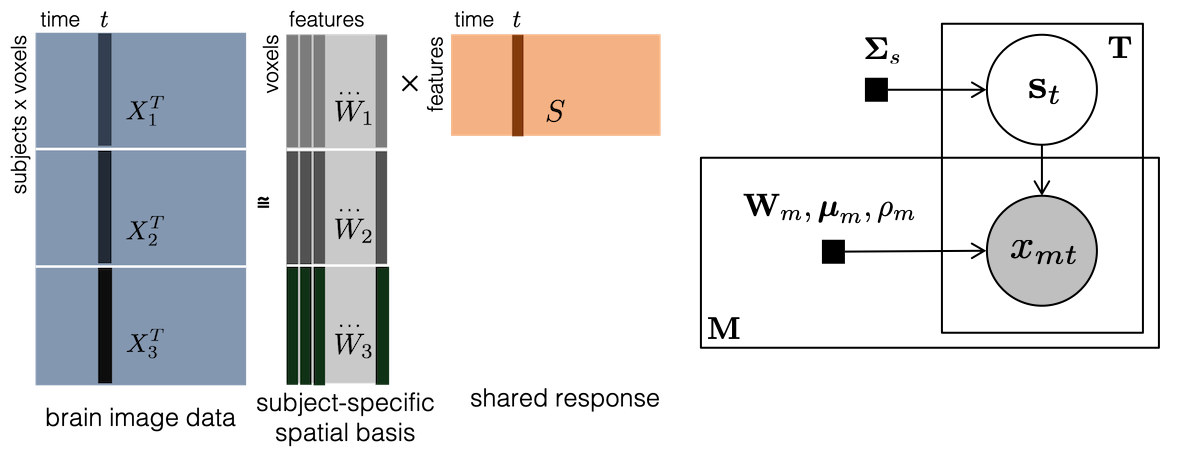}
\caption{\textbf{Left:} Illustration of deterministic SRM for three subjects. \textbf{Right:} Graphical model for SRM with $M$ subjects and $T$ time points (adapted from ~\citep{chen2015reduced}). Brain image data $x_{mt} \in \RR ^{V_m}$ ($V_m$ voxels) is observed from subject $m$ at time $t$, $t = 1:T, m = 1:M$. Each observation $x_{mt}$ is a linear combination of subject-specific orthogonal basis (columns of $\mathbf{W}_m$) using the weights specified by $s_t$. The two plates are repeated $T$ and $M$ times, respectively. Shaded nodes: observations, unshaded nodes: latent variables, and black squares: parameters.}
 \label{fig:SRM}
\end{figure}

\vspace{0.1in}
\noindent\textsl{- Translating assumptions to a graphical model: shared response as a latent variable}\\
To translate the assumptions above into a computational model, let us look at the deterministic SRM first and then the probabilistic version. The deterministic SRM factorizes the transpose of each subject's brain image data $\mathbf{X}_m^T$ into a subject-specific spatial basis $\mathbf{W}_m$ and the shared response $\mathbf{S}$ with the orthogonal constraint $\mathbf{W}_m^T \mathbf{W}_m=I$ (Fig.~\ref{fig:SRM}), where $\mathbf{X}_m \in \RR ^{T \times V_m}$ is the brain image data of subject $m$, $\mathbf{W}_m \in \RR ^{V_m \times K}$ is the subject-specific spatial basis of subject $m$, $\mathbf{S} \in \RR ^{K \times T}$ is the shared response across subjects, $V_m$ is the number of voxels of subject $m$, $T$ is the number of time points, and $K$ is the number of features (the dimensionality of the shared response). $K$ is a tunable hyper-parameter which is usually much smaller than $T$. More formally, deterministic SRM minimizes the Frobenius norm of reconstruction error  
\begin{equation}
\left \| \mathbf{X}_m^T-\mathbf{W}_m\mathbf{S} \right \|_F ^2
\end{equation}
 under the constraint $\mathbf{W}_m^T \mathbf{W}_m=I$. 
This simple model is then extended to a probabilistic setting, as shown in Fig.~\ref{fig:SRM}. Here $x_{mt} \in \RR ^{V_m}$ denotes the observed brain image data of subject $m$ at time $t$, $s_t \in \RR ^K$ denotes a shared latent random vector with 
\begin{equation}
s_t \sim \mathcal{N}(0, \Sigma _s).
\end{equation}
The distribution of $x_{mt}$ conditioned on $s_t$ is then 
\begin{equation}
x_{mt}|s_t \sim \mathcal{N}(\mathbf{W}_ms_t+\mu_m, \rho_m^2I),   
\end{equation}
where the subject-specific average $\mu_m$ accounts for non-zero mean and $\rho_m^2I$ is the subject dependent isotropic noise covariance (for non-isotropic noise covariance in SRM, see~\ref{S:mike}). In the probabilistic version, the orthogonal constraint still holds. A constrained expectation-maximization (EM) algorithm is used to solve this model.

\vspace{0.1in}
\noindent\textsl{- Applications: identified shared responses and extensions}\\
The SRM identifies the shared and individual responses in a multi-subject fMRI dataset with naturalistic tasks. The explicit structure of SRM makes the fMRI data it is applied to more interpretable: 
the extracted shared responses allow us to aggregate information from multiple subjects, and the individual responses could be used to identify what is unique for each subject. SRM shows improved performance in various tasks, such as image-viewing fMRI data classification, using shared and individual responses, as described in \cite{chen2015reduced}, and movie scene classification \cite{vodrahalli2018mapping}. 

Compared with hyperalignment (HA) \cite{haxby2011common}, SRM also has a built-in dimensionality reduction mechanism with a tunable number of features, where HA is an earlier multi-subject alignment algorithm with the objective to minimize 
\begin{equation}
\left \| \mathbf{W}_m^T\mathbf{X}_m^T-\mathbf{S} \right \|_F ^2 
\end{equation}
under the constraint $\mathbf{W}_m^T \mathbf{W}_m=I$, $\mathbf{W}_m \in \RR ^{V_m \times V_m}$. Note that $\mathbf{W}_m$ is a square matrix here because HA aims to rotate each subject's $\mathbf{X}_m$ to match a global template $S$. More importantly, if $\mathbf{W}_m$ in HA is set to $\RR ^{V_m \times K}$ as in SRM, then it sometimes learns uninformative $S$. As illustrated in \cite{chen2015reduced}, when performing an image stimulus classification experiment, HA with $\mathbf{W}_m \in \RR ^{V_m \times K}$ shows much lower testing accuracy than SRM.

Furthermore, SRM already has several extensions which make it more useful. For example, searchlight SRM~\cite{zhang2016searchlight} combines SRM with searchlight analysis, which enables the localization of shared responses. Multi-dataset multi-subject analysis (MDMS)~\cite{zhang2018transfer} extends SRM to the multi-dataset setting where the model can aggregate information across subjects and datasets with different stimuli. Semi-supervised SRM~\cite{turek2017semi} combines SRM with an additional multinomial logistic regression objective, such that the model can leverage partially labeled data. Matrix-normal SRM~\cite{Shvartsman2018}, discussed below, makes different choices in modeling the residuals and the constraints on $\mathbf{W}_m$. 
All of these recent development of SRM illustrate the expandability of PGM-based method and the flexibility of adapting one PGM to different experimental settings and research purposes. In all these developments, the effectiveness of inverting a PGM can be verified by simulation, which is difficult for algorithms that do not explicitly build PGMs.  

%% file: ivy.tex
\noindent\textsl{- Defining the problem: discovering full-brain functional connectivity}\\
Recent research suggests that the functional connectivity (networks) in human brain, commonly represented by the spatial covariance structure of fMRI data, can change during different cognitive states~\cite{Turk13}. To estimate functional connectivity during a particular cognitive state (or an experimental condition) from fMRI data, one approach is to compute the correlation between the time series of pairs of voxels~\cite{RubiSpor10}. Because of the computational time and memory demanded by this voxel-based approach, most researchers focus their analysis on pre-selected regions of interest (ROIs). But this requires anatomically predefined ROIs which may not correctly capture the voxel-wise correlation. Voxel-based methods such as independent component analysis~\cite{beckmann2005investigations,calhoun2012multisubject} generate statistically independent spatial maps and they are useful for applications that assume statistical independence between different neural sources. But each component discovered in this approach, or their combination, cannot be easily used to analyze spatially overlapping but functionally distinct activity patterns. Topographic Factor Analysis (TFA)~\cite{MannEtal14b} and Hierarchical Topographic Factor Analysis (HTFA) are Bayesian factor analysis models that can be used to efficiently analyze full-brain functional connectivity in large multi-subject neuroimaging datasets~\cite{Htfa18}. Further, one of properties of HTFA is to generate spatially compact factors that partially overlap, and this property can help analyze and detangle the contributions of activity patterns that are functionally distant but spatially overlapping.

\vspace{0.1in}
\noindent\textsl{- Making assumptions: spatial function-based latent factors}\\
Both TFA~\cite{MannEtal14b} and HTFA~\cite{Htfa18} cast each subject's brain images as a linear combination of latent factors, where each latent factor is modeled as a parameterizable spatial function. Each latent factor can be interpreted as a node in a simplified representation of the brain's network. A subject's matrix of the changing weights on the nodes over time may be viewed as a low-dimensional embedding (or representation) of the original brain data. The pairwise correlations between each factor's weights over time further reflect the signs and strengths of the node-to-node connections (i.e.\ the functional connectivity). Both TFA and HTFA approximate each subject's functional connectivity by firstly representing each brain image in terms of the activities of a set of localized network nodes, and then computing the covariance of the activity.  Furthermore, HTFA~\cite{Htfa18} is a multi-subject extension of TFA~\cite{MannEtal14b}, and attempts to discover the network nodes that are common across a group of subjects. HTFA estimates a global template as well as each individual's subject-specific template. The global template describes where each common network node is placed, how wide it is and how active it tends to be. Each subject-specific template is a particular instantiation of the common network nodes and the subject's node activities.

\begin{figure}[!t] 
   \centering
   \includegraphics[width=1.0\textwidth]{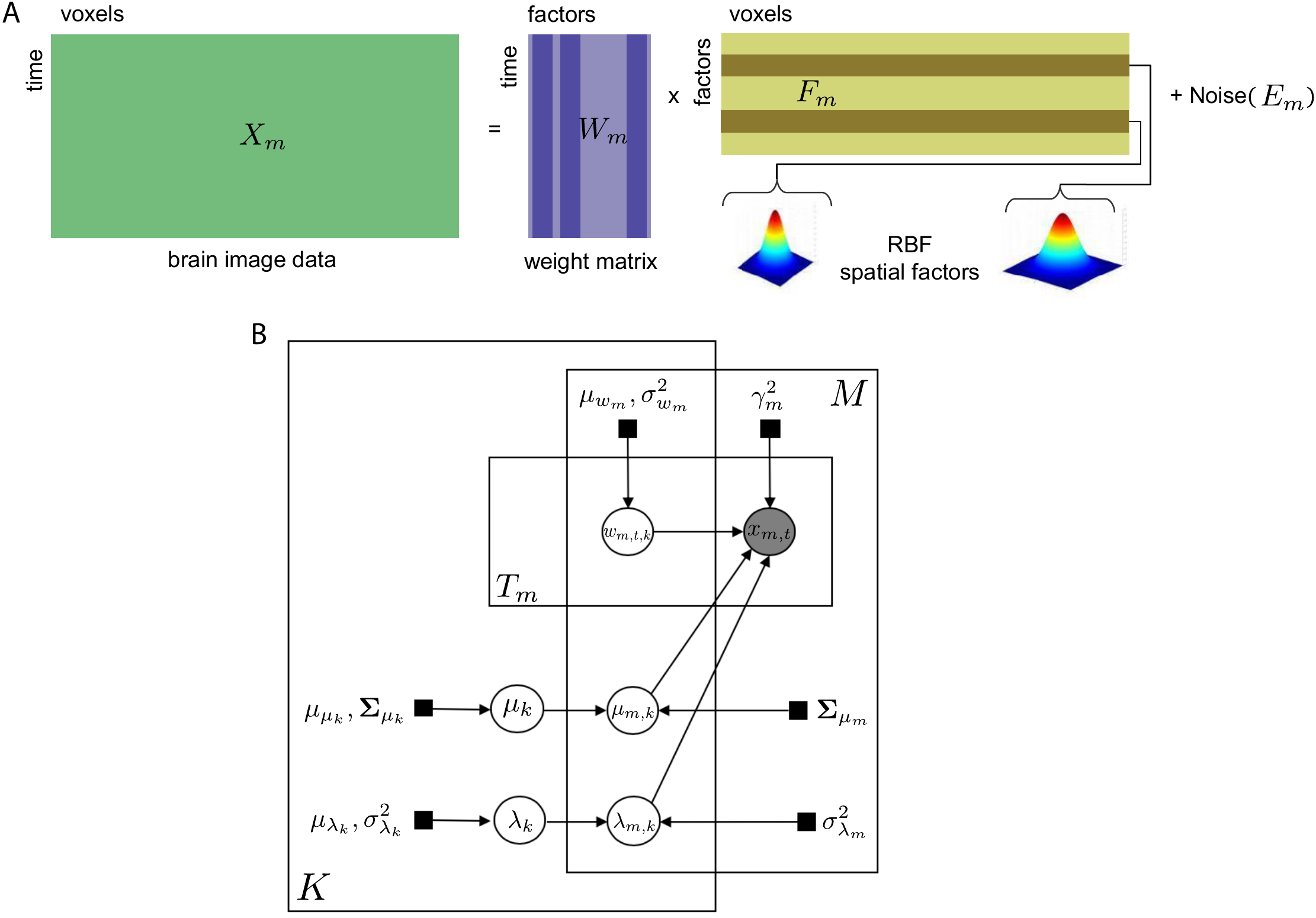} 
   \caption{A)(H)TFA factor model. fMRI data $\mats{X}{m}$  is decomposed into weight maxtrix $\mats{W}{m}$ and factor matrix $\mats{F}{m}$. Each factor is a RBF function. B) Graphical model for HTFA (adapted from ~\cite{Htfa18}). Brain image data $x_{m,t} \in \RR ^{V_m}$ ($V_m$ voxels) is observed from subject $m$ at time $t$, $t = 1:T_m, m = 1:M$. Each observation $x_{m,t}$ is a linear combination of $K$ subject-specific latent factors, using the weights specified by $w_{m,t,k}$. Each latent factor (row of $F_m$) is a spatial function of $\ves{\mu}{m,k}$ and $\lambda_{m,k}$). The three plates are repeated $K$, $T_m$ and $M$ times, respectively. Shaded nodes: observations, unshaded nodes: latent variables, and black squares: parameters.}
   \label{fig:htfa_gm}
\end{figure}

\vspace{0.1in}

\noindent\textsl{- Translating assumptions to a graphical model: global and subject specific template}\\
HTFA is formulated as a probabilistic latent variable model. Let $\mats{X}{m} \in \real{T_m \times V_m}$ represent subject $m$'s data as a matrix with $T_m$ fMRI samples of the activity of $V_m$ voxels, each sample being vectorized as one row in \mats{X}{m}.
Then, each subject is approximated with a factor analysis model
\begin{equation}
\mats{X}{m} = \mats{W}{m} \mats{F}{m} + \mats{E}{m}, \label{eq:htfa_factor_analysis}
\end{equation}
where $\mats{W}{m} \in \real{T_m \times K}$ are the weights of $\mats{F}{m} \in \real{K \times V_m}$, the latent factors. \mats{E}{m} is the noise term. Each latent factor (row of $\mats{F}{m}$) is a radial basis function (RBF) with center at \ves{\mu}{m,k} and width $\lambda_{m,k}$

\begin{equation}
f_{v,m,k}\left(\ves{p}{v};\ves{\mu}{m,k},\lambda_{m,k}\right) = \exp \left\{-\frac{ \normtwo{\ves{p}{v}-\ves{\mu}{m,k}}^2}{\lambda_{m,k}} \right\},
\label{eq:htfa_factor}
\end{equation} 
in positions $\ves{p}{v} \in \real{3}$ for all the voxels in the three-dimensional voxel space of the brain. HTFA defines the local factors in \mats{F}{m} as perturbations of the factors of a global template in \mat{F}. 
Therefore, the factor centers $\ves{\mu}{m,k}$ for all subjects are obtained from a multivariate normal distribution with mean \ves{\mu}{k} and covariance \mats{\Sigma}{\ves{\mu}{m}}. The mean \ves{\mu}{k} represents the center of the global $k^{th}$ factor, while \mats{\Sigma}{\ves{\mu}{m}} determines the distribution of the possible distance between the global and the local center of the factor. Similarly, the widths $\lambda_{m,k}$ for all subjects are drawn from a normal distribution with mean $\lambda_k$, the width of the global $k^{th}$ factor, and variance $\sigma_{\lambda_m}^2$. The model defines multivariate Gaussian prior \normaldist{\ves{\mu}{\mu_k}}{\mats{\Sigma}{\mu_k}} for the global parameters \ves{\mu}{k} and Gaussian prior \normaldist{\mu_{\lambda_k}}{\sigma_{\lambda_k}^2} for $\lambda_k$, respectively. 
In addition, the columns of the weight matrices \mats{W}{m} are modeled with a \normaldist{\mu_{w_m}}{\sigma_{w_m}^2} distribution and the elements in the noise term \mats{E}{m} are assumed to be independent with a  \normaldist{0}{\gamma_{m}^2} distribution (for one approach to non-independent noise, see~\ref{S:mike}). The associated graphical model is shown in Fig.~\ref{fig:htfa_gm}. 

\vspace{0.1in}

\noindent\textsl{- Solving the model}\\
The \textit{maximum a posteriori} (MAP) probability estimation procedure is used to solve the HTFA model. The method consists of a global and local step that iteratively update the parameters~\cite{BigData2016}. The global step updates the parameters of the $K$ distributions in the global template. The local step updates for each subject $m$ the weight matrices $\mats{W}{m}$, the local centers $\ves{\mu}{m,k}$ and the widths $\lambda_{m,k}$ of each latent factor. To update the parameters of the factors in \mats{F}{m}, the local step solves the following problem, where $\phi_m$ is a subsampling coefficient. Optimized implementations of TFA and HTFA~\cite{BigData2016} can be found in BrainIAK~\cite{kumar2019brainiak}.

\begin{align}
\left\{\vehs{\mu}{m,k},\hat{\lambda}_{m,k} \right\}_{k} = 
 \argmin_{\left\{\ves{\mu}{m,k},\lambda_{m,k} \right\}_k} [\frac{1}{2{\gamma}_m^2}\normfro{\mats{X}{m} - \mats{W}{m}\mats{F}{m}}^2  \nonumber\\
+\frac{1}{2\phi_m} \sum_{k=1}^{K} \left(\ves{\mu}{m,k} - \vehs{\mu}{k}\right)\mats{\Sigma}{\mu_m}^{-1} \left(\ves{\mu}{m,k} - \vehs{\mu}{k}\right)^T \nonumber\\
+ \frac{1}{2\phi_m \sigma_{\lambda_m}^2 }\sum_{k=1}^{K} \left(\lambda_{m,k} - \hat{\lambda}_{k} \right)^2 ]
\label{eq:htfa_subject_update_rbfs}
\end{align} 
 Eq.~\eqref{eq:htfa_subject_update_rbfs} consists of the reconstruction error, the Mahalanobis distance between global and local centers, and the Euclidean distance between global and local widths.  Due to its non-linearity, the latent factors of each subject are computed using a non-linear least squares solver~\cite{kumar2019brainiak}, and implemented with a trust-region reflective method~\cite{Coleman1996trf}. The weight matrix is solved with a closed-form solution of the form of ridge regression. The hyper-parameters of the global template are updated given the local estimates and under the assumption that the posterior has a conjugate prior with multivariate normal and normal distribution for centers and width, respectively.
\vspace{0.1in}

\noindent\textsl{- Advantages}\\
Because the number of network nodes is typically substantially smaller than the number of fMRI voxels, one obvious advantage of HTFA is that it can be orders of magnitude more efficient than traditional voxel-based functional connectivity approaches. Compared to other dimensionality reduction methods, HTFA provides additional advantages: (a) it provides estimation of both global and subject-specific templates, and builds connections between them; (b) modeling the latent factors as spatially smooth allows them to be overlapping rather than distinct, as would be the case of functional connectivity based on anatomically defined brain region segmentation; (c) it provides a natural means of determining how many network nodes (latent factors) should be used for a given dataset (further details about determining $K$ can be acquired from ~\cite{Htfa18});  and (d) because HTFA decomposes brain images into sums of spatial functions, it supports seamless mapping between images of different resolutions and potentially different imaging modalities.
\vspace{0.1in}

\noindent\textsl{- Applications}\\
HTFA can be applied to different tasks with multi-subject fMRI datasets, for example, inferring dynamic full-brain inter-subject functional connectivity when participants are listening to a story or watching a television show~\cite{Htfa18}. The functional connectivity of each subject can be estimated by the correlation between the column of \mats{W}{m}.  Since the global template of HTFA makes sure the columns of the \mats{W}{1...M} correspond to the same network nodes across the different subjects, the ISFC can be computed by the correlation between the columns of \mats{W}{1...M} across subjects. A recent study showed both HTFA-derived activities and HTFA-derived ISFC can be used to reliably decode which moments in the story or show the participants were experiencing. A decoder with the combination of these two types of patterns outperformed decoders with either activity or connectivity patterns alone~\cite{Htfa18}.

%% file: anqi.tex
\vspace{0.1in}
\noindent\textsl{- Defining the problem: fMRI decoding with sparse weights}\\

A primary research problem neuroscientists have been studying with fMRI is brain decoding or inverse inference \citep{o2007theoretical,laconte2005support,cox2003functional}. The goal of a decoding task is to understand how brain activity represents task-related variables, e.g.\ the orientation of a grating \cite{haynes2005predicting} or the category of an object \cite{HaxbyGobbiniFureyEtAl2001}. Researchers often use linear classification and regression methods to identify the brain regions or voxels that are most closely related to these task-related variables by inspecting the decoding weights.

A piece of domain knowledge in fMRI decoding is that different regions of the brain are specialized for different functions, implying that only few small regions of the brain are specifically activated during an individual task. In the linear regression methods that are common in the field, this assumption is equivalent to assuming that the weights mapping fMRI to task-related variables are mostly zeros with a few non-zero values, which is referred to as ``sparsity''. This model assumption is also reasonable from a statistical standpoint, since the task variable is linked to fMRI data with usually tens of thousands of voxels, but the number of fMRI volumes with valid task labels is far smaller, e.g.\ a few hundred. We need to estimate tens of thousands of coefficients to map a full brain pattern down to a single task variable given only a few hundred observations. This is referred to as a high-dimensional and small-sample issue, where the linear regression model would fit seemingly predictive information from noise instead of the underlying brain signal, and thus would not generalize well to new data. To address this issue, one can reduce the number of coefficients. With the sparsity assumption, we effectively regularize the linear decoding model by restricting the weight parameter space to a much smaller one, thus mitigating the issue. 

\vspace{0.1in}
\noindent\textsl{- Making assumptions: region sparsity}\\
Sparse decoding has already been exploited in the previous literature \citep{carroll2009prediction,michel2011total,grosenick2013interpretable}. However, the non-zero coefficients are not randomly distributed throughout the brain, but tend to arise in clusters, and are therefore not independent a priori. Sets of voxels allowing to discriminate between different brain states are expected to form small localized and connected areas. If one voxel encodes information related to the task, its neighboring voxels should carry similar information, given that contiguous brain regions of shared functions extend over multiple adjacent voxels. This type of sparsity is referred to as ``region sparsity'' \citep{wu2019dependent}. By considering such region sparsity, one can impose a structured sparsity regularization over the decoding weights which further constrains the parameter space to search and thus eases the decoding weights optimization task. Wu et al.\ \citep{wu2019dependent} developed a Bayesian framework that incorporated such region sparsity into brain decoding for fMRI analysis and showed superior decoding performance and more biologically informed decoding weights for three brain imaging datasets. 

\vspace{0.1in}
\noindent\textsl{- Translating assumptions to a graphical model: building a region sparsity prior over the brain weights}\\
The model proposed in \citep{wu2019dependent} is referred to as  ``Dependent Relevance Determination'' (DRD). It builds a Bayesian hierarchical model that imposes a sparsity prior over the decoding weights. Unlike previous work with sparsity assumptions, DRD also assumes that nearby sparse voxel-activations should be correlated to each other based on their spatial locations. 

Formally, the fMRI decoding problem can be formulated in a linear regression setting: at time $t$, consider a scalar response $y_t \in \RR$ linked to an input vector
$\vx_t \in {\RR}^V$ via the linear model:
\begin{equation}\label{linear}
y_t = \vx_t^\top \vw + \epsilon_t , \quad \mbox{for} \quad t = 1, 2, \ldots, T,
\end{equation} 
with observation noise  $\epsilon_t \sim \Nrm(0,
\nsevar)$, where $T$ is the number of time points and $V$ is the number of voxels.
The regression (linear weight) vector $\vw \in \RR^V$ is
the quantity of interest. We can denote the fMRI data matrix by $\mathbf{X} \in \RR^{T
  \times V}$, where each row of $\mathbf{X}$ is the $t^{th}$ input vector
$\vx_t\trp$ and $T\ll V$, and the observation vector by $\vy = [y_1, \cdots,
y_T]\trp\in \RR^{T}$. Since the noise is Gaussian, it can be written as
\begin{equation}
\vy|\mathbf{X}, \vw, \nsevar \sim \Nrm(\vy|\mathbf{X} \vw, \nsevar \mathbf{I}).
\end{equation} 
DRD imposes a zero-mean multivariate normal prior on $\vw$:
\begin{equation}\label{eq:C}
\vw |\vtheta \sim  \mathcal{N}(0, C(\vtheta)),
\end{equation} where the prior covariance matrix $C(\vtheta)$ is a
function of hyperparameters $\vtheta$. One can specify $C(\vtheta)$
based on prior knowledge on the regression vector, e.g. sparsity
\cite{tipping2001sparse, tipping2002analysis, Wipf08}, smoothness
\cite{sahani2003evidence, schmolck2008smooth}, or both
\cite{park2011receptive}.
Ridge regression assumes $C(\theta)=\theta^{-1}I$ where $\theta$ is a scalar for precision and $I$ is the identity matrix. Automatic relevance determination (ARD) \cite{neal2012bayesian} uses a diagonal prior covariance matrix with a distinct hyperparameter $\theta_i$ for each element of the diagonal, thus $C_{ii}=\theta_i^{-1}$. DRD is an extension of ARD by imposing dependency between $\theta_i$. 

Given the general Bayesian linear regression setting, DRD aims to construct a covariance $C(\vtheta)$ which generates the region-sparse $\vw$. This is achieved by introducing a latent variable $\vu \in \RR^{V}$. $\vu$ is from a Gaussian process (GP) prior, i.e.\
\begin{equation}\label{gp}
  \vu \sim \Nrm(b\vones, k).
\end{equation}  
A Gaussian process \citep{rasmussen2003gaussian} is a stochastic process whose realizations are draws from a multivariate normal distribution, but whose mean $b$ and covariance $k$ can be functions of another input (e.g. spatial locations). For example, by defining a Gaussian process with covariance (kernel) that is a function of spatial distances, we can constrain the samples drawn from the Gaussian process distribution to exhibit spatial correlation based on the kernel. Most commonly in GPs, the squared exponential kernel is used, which constrains the draws from the multivariate normal to be smooth over space, i.e.\ $k(\chi,\chi')=\rho\;\mbox{exp}(-\frac{||\chi-\chi'||^2}{2l^2})$ where $\chi$ and $\chi'$ are the spatial location of any two voxel. Functions sampled from such a GP are smooth functions. The smoothness is determined by the length scale $l\in \RR$ and the magnitude of the functions is determined by $\rho \in \RR$. These three hyperparameters in the DRD prior are jointly denoted by $\vtheta = \{ b, \rho, l \}$. 

By imposing a GP prior over the latent $\vu$, DRD effectively captures dependencies in $\vu$. Given such latent, Wu et al.\ formulate the covariance of $\vw$ with
\begin{equation}\label{exp_non}
C = {\mbox{diag}}[\exp(\vu)].
\end{equation} 
The exponential function here ensures the non-negativity of values on the diagonal of $C$, which makes it a valid covariance. When the mean $b$ is very negative, $\exp(\vu)$ has many close-to-zero values that result in soft-sparsity (since their prior mean is zero and the variance is nearly zero as well). Note that the spatial smoothness of $\vu$ induces dependencies between the variances of nearby voxels, that is, the prior variance changes slowly between neighboring
coefficients.  If the $i^{th}$ coefficient of $\vu$ has a large prior
variance, then probably the coefficients of its adjacent voxels are large as
well.

\begin{figure}[!ht] 
   \centering
   \includegraphics[width=\linewidth]{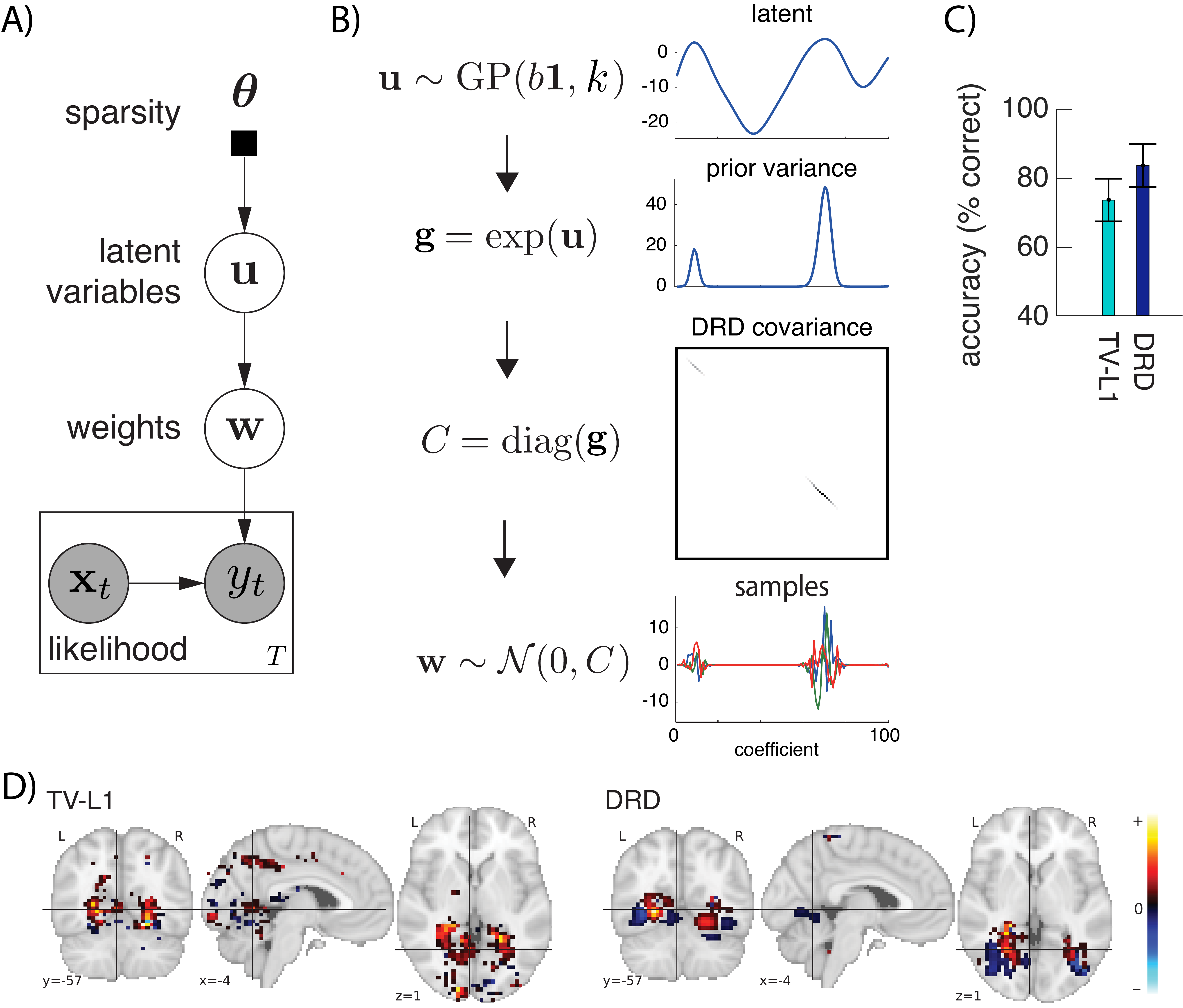} 
   \caption{A) Probabilistic graphical model for DRD. The rectangular box indicates a graph for each time point. Each fMRI volume $\vx_t$ at time $t$ is mapped to the experimental response $y_t$ together with a global variable $\vw$ (eq.~\ref{linear}). The decoding weight vector $\vw$ is conditioned on a latent variable $\vu$ (eq.~\ref{eq:C} and \ref{exp_non}). The latent variable $\vu$ is generated from some hyperparameters in $\vtheta$ (eq.~\ref{gp}). B) The generating process for region-sparse decoding weight $\vw$. C) Accuracy comparison between TV-L1 and DRD. The accuracy value is averaged over all pairs of objects. D) Decoding weight map for the house vs bottle pair using TV-L1 (left) and DRD (right). Yellow indicates very positive values and light blue indicates very negative values. Black means small values. (adapted from ~\citep{wu2019dependent})}
\label{drd}
\end{figure}

Fig.~\ref{drd}A and B show the probabilistic graphical model of DRD and the process to generate region-sparse samples for $\vw$. 

\vspace{0.1in}
\noindent\textsl{- Solving the model}\\
In the paragraphs above, we show how to build a generative model for DRD to generate region-sparse decoding weights. When using DRD, one can apply it to fMRI decoding problems where we have the imaging data $\mathbf{X}$ and prediction targets $\vy$, and we aim to infer the decoding weight vector $\vw$. To solve this problem, we need to reverse the generating process using some inference methods. Exact Bayesian inference is infeasible with a DRD prior. However, approximate inference can be carried out efficiently using both Laplace approximation and Markov Chain Monte Carlo (MCMC) sampling. Further details regarding inference can be acquired from \citep{wu2019dependent}.

\vspace{0.1in}
\noindent\textsl{- Application: classification on a visual recognition task}\\
The visual recognition dataset \cite{haxby2001distributed} is from a study on object representation in human ventral temporal cortex. In the object recognition experiment, 6 subjects were asked to recognize 8 different types of objects (bottles, houses, cats, scissors, chairs, faces, shoes and scrambled control images). Wu et al.\ \citep{wu2019dependent} examined this dataset to learn the weights mapping the fMRI brain activity to object categories for each subject. They cast the multi-category classification problems into multiple binary classification problems for each pair of categories. Wu et al.\ employed the same linear regression model as in eq.~\ref{linear} for training the model. When making predictions, they took the sign of the output $y$ as the discrete binary labels ($+1/-1$).

They showed that DRD achieved the highest accuracies for most of the binary classifications compared with other state-of-art sparse decoding methods \citep{michel2011total,grosenick2013interpretable}. Fig.~\ref{drd}C shows a comparison of accuracies between DRD and a baseline model, total variation L1 (TV-L1) \cite{gramfort2013identifying}. More specifically, DRD is able to find more biologically informed decoding weight maps for many pairs compared with TV-L1. By saying biologically informed, we mean that only small regions of voxels are correlated to a specific task and nearby voxels are more likely to be activated together compared with LASSO. Fig.~\ref{drd}D presents the brain map estimation for the house-vs-bottle pair for TV-L1 and DRD. DRD weights have significant positive regions in the parahippocampal place area (PPA) (responding more strongly to scenes depicting places) \cite{epstein1999parahippocampal} and clustered negative weights in the lateral occipital complex (LOC) (responding to objects in human occipito-temporal cortex) \cite{eger2008fmri}. By comparison, TV-L1 weights in LOC are not very clustered and don't show negative activations. 

We describe the DRD model here in a generative way. The brain decoding weights are generated from a DRD prior, but the application is a discriminative model, i.e.\ mapping fMRI data to experimental variables. Because the DRD prior was proposed to learn region-sparse brain weights regardless of whether a model is discriminative or generative, it can also be inserted to generative models such as the factor analysis models in SRM in essentially the same way.

%% file: mingbo.tex
\vspace{0.1in}
\noindent\textsl{- Defining the problem: neural pattern similarity} \\
As sensory inputs get processed in the brain, each neural population of one brain region performs nonlinear computation of the input from neurons of other regions. The representation of the same external object thus changes from one region to another. One fundamental question in neuroscience is how these representations are transformed, in service for deciding the right actions to take \cite{dicarlo2012does,marr1978representation}. One way to describe representation is in terms of what stimuli are encoded closer and what are encoded farther apart. Beyond studying representation of external stimuli, the same question can also be asked about different cognitive states: which states are represented closer in a brain region? 


Early behavioral studies investigated representations of objects by asking people to judge how similar a pair of stimuli are to each other \cite{shepard1970second}. The structure of the similarity matrix, composed of the judged degrees of similarity between all pairs of tested stimuli, reflects the geometry of the internal representational space being used to encode stimuli. Such approach is limited to representations accessible for conscious report \cite{ericsson1980verbal}.
To overcome this and to compare computational models against multiple types of neural data, Kriegeskorte et al.\ \cite{KriegeskorteMurBandettini2008} proposed Representational Similarity Analysis (RSA), which utilizes neural recordings to understand the structure of representations. This analysis assumes that the similarity between the neural patterns elicited by each pair of stimuli in a brain region reflects the similarity between the representations of these stimuli in that region. Because it does not rely on subjective judgment, RSA can be applied to studying representation in any stage of sensory processing \cite{ConnollyGuntupalliGorsEtAl2012,iordan2015basic}. Measuring similarity between neural activity patterns evoked by sensory stimuli or cognitive states is its central goal. 

\vspace{0.1in}
\noindent\textsl{- Making assumptions: relations of representational structure, neural patterns and fMRI data} \\
In order to infer the similarity between neural activity patterns, one needs to first make assumptions about the relations between the neural patterns and the recorded neural data, and between the similarity structure and the patterns. 

\begin{figure}[!ht] 
   \centering
   \includegraphics[width=1.0\textwidth]{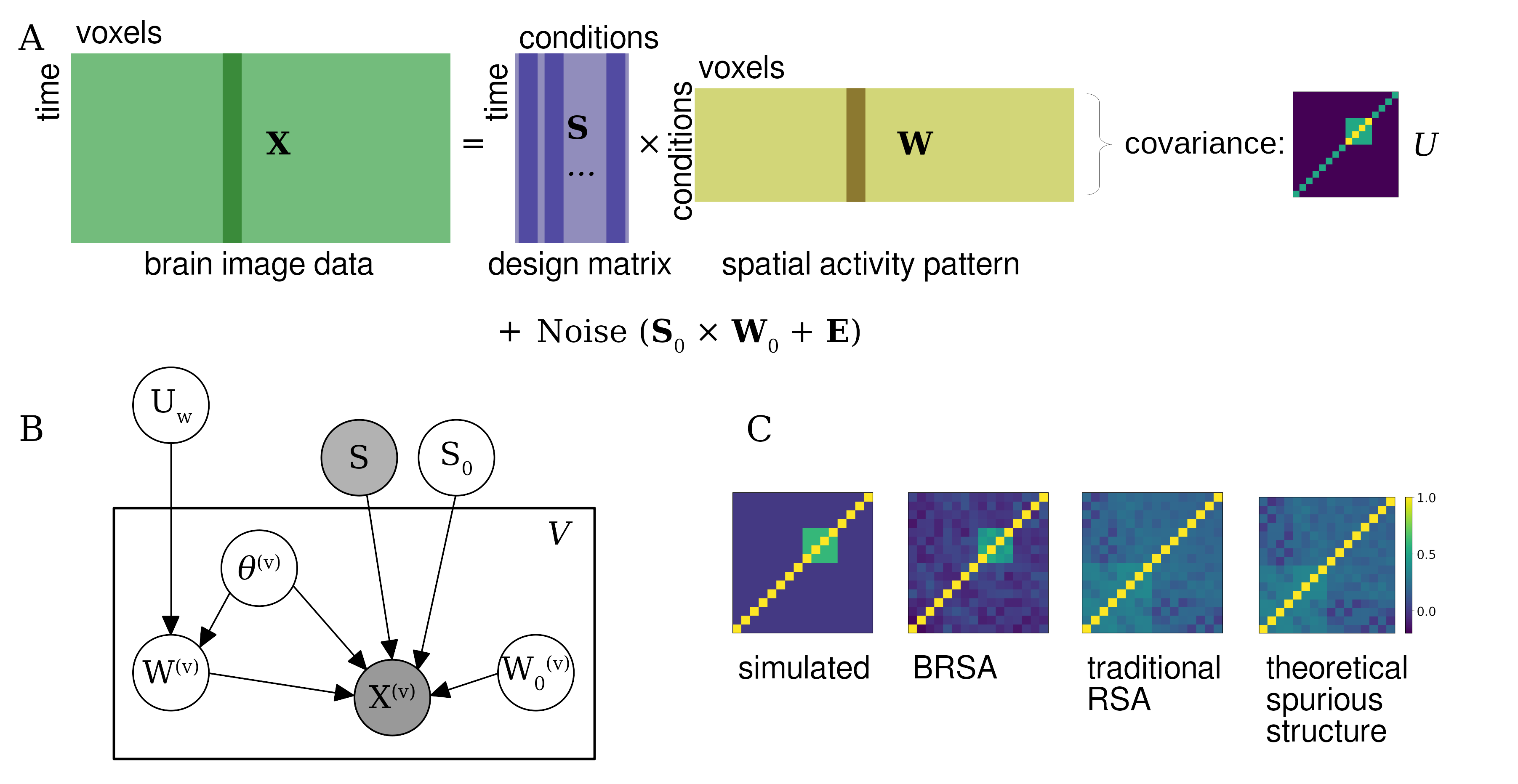} 
   \caption{A) BRSA assumes a similar factor model as SRM and (H)TFA. To capture both spatial and temporal correlation in residual noise, the noise is further modeled by a factor decomposition of spatially correlated noise plus spatially independent noise. Additionally, each column of the weight matrix $\mat{W}$ (activation patterns) are assumed to share the same covariance structure, which underlies the similarity between patterns. B) Probabilistic graphical model for BRSA. The rectangular plate is repeated for each voxel. Variables within the plate are voxel-specific and those outside the plate are shared by all voxels. $U_W$ is the target to estimate but is indirectly related to $\mat{X}$ through unknown patterns $\mat{W}$. To infer $U_W$, other unknown variables are either marginalized or (in the case of $\mat{S}_0$) determined through an iterative fitting procedure (see \cite{cai2019representational}) C) The simulated similarity structure, the similarity structures recovered by BRSA, by correlation of point estimates of $\mat{W}$ (data-mining approach) and the theoretical spurious structure expected to be introduced by the design matrix $\mat{S}$ when estimating $\mat{\hat{W}}$. B) and C) are adapted from \cite{cai2019representational} }
\label{fig:brsa}
\end{figure}

The neural activity of a region\footnote{RSA typically focuses on single brain region instead of the whole brain.} during a task can be considered as being generated by the sum of various spatial patterns, each being modulated by different time courses. In this sense, the basic assumption of fMRI data underlying RSA is also a factor model, as in SRM and (H)TFA (Fig.~\ref{fig:brsa}A).
 The difference here is that at least a subset of the modulation time courses are explicitly tied to when and how much the brain is engaged in each task condition, which are pre-defined by the researchers. The spatial pattern being modulated by each time course is the relative degree by which different voxels are activated by the task condition. In addition to the activity explained by the temporal modulation of these patterns, the data also contain unexplained fluctuation with both spatial and temporal correlation. Therefore, the similarity matrix one seeks to estimate is only indirectly related to the noisy fMRI data through unknown neural activity patterns and their modulation time courses predicted by the task. 
 
 There are many ways to define similarity. One way is based on the cosine of the angle between the vectors corresponding to activity patterns in the space spanned by the voxel activation levels, which is adopted by the algorithm of Bayesian RSA (BRSA) \cite{cai2019representational,cai2016bayesian}. Other common ways include correlation between demeaned patterns, and Euclidean distance or Mahalanobis distance between patterns (as measures of dissimilarity) \cite{KriegeskorteMurBandettini2008,diedrichsen2016distribution,ramirez2017representational,nili2014toolbox}. Here we focus on cosine of angle between patterns, which can be alternatively considered as correlation without demeaning patterns.

\vspace{0.1in}
\noindent\textsl{- Translating assumptions to a graphical model: two-stage model of fMRI data with representational structure as latent variable}\\

Since the time course of a task is known, the modulation time course (so-called design matrix) can be constructed based on the timing of the task conditions and the shape of the smooth delayed response (the haemodynamic response function, HRF) in fMRI signals following neuronal activity. We denote the design matrix as $\mat{S} \in \RR ^{T \times K}$, where $T$ is the total time points and $K$ is the number of task conditions in an experiment. Then, the factor model of fMRI data can be expressed as
\begin{equation}
    \mat{X} =  \mat{S} \mat{W} + \mats{S}{0} \mats{W}{0} + \mat{E}.
\label{brsa_voxel}
\end{equation}
Here, $\mat{X} \in \RR ^ {T \times V}$ is the time by voxel matrix of the fMRI time series in a region of interest, where $V$ is the total number of voxels in that region. $\mat{W} \in \RR ^ {K \times V}$ is the unknown activation patterns associated with all the task conditions. $\mats{S}{0} \mats{W}{0}$ captures spatially correlated fluctuation unrelated to the task. $\mat{E}$ denotes the residual spatially independent noise, but it can have temporal autocorrelation, which may be modeled with an auto-regressive (AR) process such as AR(1) (for an alternate approach to the residual noise in RSA, see~\ref{S:mike}). Generally, researchers do not have full knowledge of $\mats{S}{0}$ or $\mats{W}{0}$, but may have regressors (such as the head motion time course) which accounts for some variance in $\mats{S}{0}$. 
Assuming that $\mat{E}$ is random variable drawn from the noise distribution, Eq.~\ref{brsa_voxel} implicitly defines the conditional probability of the data in each voxel given $\mat{S}$, $\mat{W}$, $\mats{S}{0}$, $\mats{W}{0}$ and the parameters $\mat{\vtheta}$ of the AR process, i.e.\ $p(X^{(v)} | \mat{S}, W^{(v)}, \mat{S}_{0},  W_{0}^{(v)}, \theta^{(v)})$ for voxel $v$.    

When cosine angle $\alpha_{i,j}$ is used as a measure of similarity between patterns $w_i$ and $w_j$ (row vectors of $\mat{W}$), $\mathrm{cos} \alpha_{i,j} = \frac{w_i w_j^T} {\sqrt{w_i w_i^T} \sqrt{w_j w_j^T} } $. If the activation profile of each voxel $w^{(v)}$ is a sample from a multivariate distribution, then $\mathop{\mathbb{E}} [w_i w_j^T]$ is the covariance between the dimensions $i$ and $j$ of this distribution \cite{diedrichsen2017representational}. By estimating the covariance structure $U_W$ of $\mat{W}$, one can obtain the cosine angle between patterns as a similarity measure. Therefore, the relation between unknown neural patterns and their similarity is modeled by assuming that each column of $\mat{W}$ is a sample drawn from a multivariate distribution with its covariance matrix being $U_W$:
\begin{equation}
    w^{(v)} \sim N(0, U_W)
\label{brsa_cov}
\end{equation}
This specifies the form of conditional probability of $w^{(v)}$ given $U_W$: $p(w^{(v)} | U_W)$. The two-stage generative model from covariance structure through activity patterns to fMRI data is depicted in Fig.~\ref{fig:brsa}B.

\vspace{0.1in}
\noindent\textsl{- Solving the model: inferring covariance structure of unknown neural patterns directly from data}\\
After the probabilistic graphical model is built and the conditional probability distribution corresponding to each edge in the graphical model is specified,
one can derive the likelihood $p(\mat{X}|U_W)$. This can be achieved by marginalizing the intermediate variables such as $\mat{W}$ and other unknown quantities that $\mat{X}$'s distribution is conditioned on ($\mat{S}_0$ is determined through an iterative fitting procedure as in \cite{cai2019representational}). Marginalization in probability refers to removing a variable in the expression of probability density by integrating the joint or conditional distribution over the variable, an important procedure in applying Bayesian models that allows the analyst to remain agnostic about the value of `nuisance' variables unimportant to the main analysis. For example, for any two variables A and B, $p(A) = \int P(A,B) dB = \int P(A|B) p(B) dB$. For any three variables A, B and C, $p(A|C)=\int p(A|B,C) p(B|C) dB$. In our case, A, B and C can be replaced by $\mat{X}$, $\mat{W}$ and $U_W$: we are agnostic about the specific mapping of the design matrix to the measurements, we are only interested in its implied covariance. In practice, the integration over several unknown variables have closed-form solution due to the assumption of Gaussian distributions in both $p(x^{(v)} | \mat{S}, w^{(v)}, \mat{S}_{0},  w_{0}^{(v)}, \theta^{(v)})$ and $p(w^{(v)} | U_W)$, which makes the computation simple. Other unknown variables can be marginalized by numerical approximation. After obtaining the formula of the marginal likelihood $p(\mat{X}|U_W)$, maximizing its logarithm with respect to $U_W$ yields the maximum likelihood estimation $\widehat{U_W}$ of $U_W$. Finally, the consine angles between $\mat{W}$ can be obtained as the correlation matrix corresponding to the covariance matrix $\widehat{U_W}$.


\vspace{0.1in}
\noindent\textsl{- Application: reducing spurious similarity structure}\\
Maximizing the likelihood $p(\mat{X}|U_W)$ while marginalizing unknown intermediate variables and uninteresting variables is a principled approach to infer the latent variable $U_W$ based on PGM. An alternative non-PGM approach is to instead first calculate $\mat{\hat{W}}$ as estimates of the unknown patterns $\mat{W}$ from the data by regressing $\mat{X}$ against $\mat{S}$, and then calculate the similarity among rows of $\mat{\hat{W}}$. This approach, however, has been shown \cite{AlinkWaltherKrugliakEtAl2015,cai2016bayesian,cai2019representational,HenrikssonKhaligh-RazaviKayEtAl2015} to introduce spurious similarity structure unrelated to the neural activity corresponding to the task of interest. 
 The reason is that although the regression provides unbiased estimates $\mat{\hat{W}}$ of the neural patterns, the covariance of $\mat{\hat{W}}$ is not the same as the covariance of $\mat{W}$: $\mat{\hat{W}}$ is contaminated by noise with specific covariance structure introduced by the regression procedure. The noise itself originates from the task-unrelated fluctuation in fMRI data. The regression procedure, at the same time of disentangling $\mat{W}$ from $\mat{X}$, also ``entangles'' the noise into each row of $\mat{\hat{W}}$ in a way that depends on the correlational structure between different columns of $\mat{S}$. The covariance structure of the noise in $\mat{\hat{W}}$ can dominate the estimated similarity structure when signal-to-noise ratio is low \cite{cai2016bayesian,cai2019representational} (Fig.~\ref{fig:brsa}C). 
 BRSA takes into account both the property of noise and uncertainty of intermediate variables $\mat{W}$, thus avoiding analyzing $\mat{\hat{W}}$ with structured noise. 

Instead of directly inferring $U_W$ from $\mat{X}$, one can alternatively assume that $U_W$ is composed of the sum of a few theoretically-motivated candidate covariance structures, and estimate the mixture coefficient of each component covariance structure. This method is called Pattern Component Modeling (PCM) \cite{DiedrichsenRidgwayFristonEtAl2011, diedrichsen2017pattern}. One can even impose a hyperprior on the the mixing coefficients, and use variational Bayesian technique to infer them \cite{friston2019variational}. The introduction of a hyperprior can incorporate additional prior assumptions or knowledge of the data. Although not directly aimed at reducing statistical bias, these methods are both developed based on clear PGMs. It is worth pointing out that in using these methods, in order to overcome the spurious similarity structure introduced by the design matrix $\mat{S}$, one still needs to either directly model the data $\mat{X}$ as in BRSA, or to model $\mat{\hat{W}}$ while explicitly taking into account the structure of the non-independent noise it carries.

Even if one takes an approach without explicitly relying on a PGM, a correct understanding of the confounding effect of noise by analyzing a PGM is helpful for developing a better non-Bayesian algorithm. For example, one can still approximate the covariance or distance structure based on the noisy patterns estimated from separate runs of experiment \cite{AlinkWaltherKrugliakEtAl2015,diedrichsen2016distribution}. This is because the noises in the patterns estimated separately from different fMRI runs come from independent sources and have zero correlation. Therefore, the covariance between estimated patterns from separate runs is an unbiased estimation of the covariance of the true unknown patterns. However, it is worth pointing out that the correlation derived from such cross-run covariance is still biased, because to calculate correlation we have to divide the covariance by the estimated standard deviations of each pattern across voxels, which is inflated by the existence of noise. To see these, one needs to understand how data $\mat{X}$ is generated from $\mat{W}$ (\ref{brsa_voxel}) and how the noise in this data generating process impacts $\mat{\hat{W}}$. Therefore, regardless of whether a researcher directly uses an analysis tool based on PGM, analyzing the data generating process and the interactions between the analysis procedures and noise is always important for realizing and avoiding any unintended consequence introduced by the chosen analysis methods.

%% file: mike.tex
\noindent\textsl{- Defining the problem: modeling spatiotemporal residuals in fMRI data}

fMRI data has structure in both the spatial and temporal dimension, and this spatiotemporal consistency needs to be exploited (or at least, managed) in order to contend with this high-dimensional and noisy data. This spatiotemporal structure exists both in the neural components corresponding to the effects of interest, and in the \emph{residual} components corresponding to everything else going on. 
In the context of supervised regression models for fMRI, practitioners tend to worry about temporal structure in both signal (by convolving the predictors with a synthetic haemodynamic response function) and residual (by performing generalized least squares, or GLS, estimation wherein the temporal structure of the residuals is modeled,~\citep[e.g.][]{Mumford2006}). More recent factor-analytic unsupervised approaches likewise assume the signal of interest itself is spatially or temporally structured due to their low-rank structure, for example the case of TFA (above) modeling brain networks as a linear combination (in time) of spatially contiguous factors. Most of these methods leave handling the residual to preprocessing stages, but this is not the only possible choice -- another is to model the spatial or temporal structure in the residuals explicity. One example of this is modeling the residual temporal autocorrelation per-voxel, as in the case of BRSA. Another, and the one we discuss here, is modeling spatiotemporable separable residuals. 

\vspace{0.1in}

\noindent\textsl{- Making assumptions: structured, separable residuals}

\begin{figure}[t]
    \includegraphics[width=\linewidth]{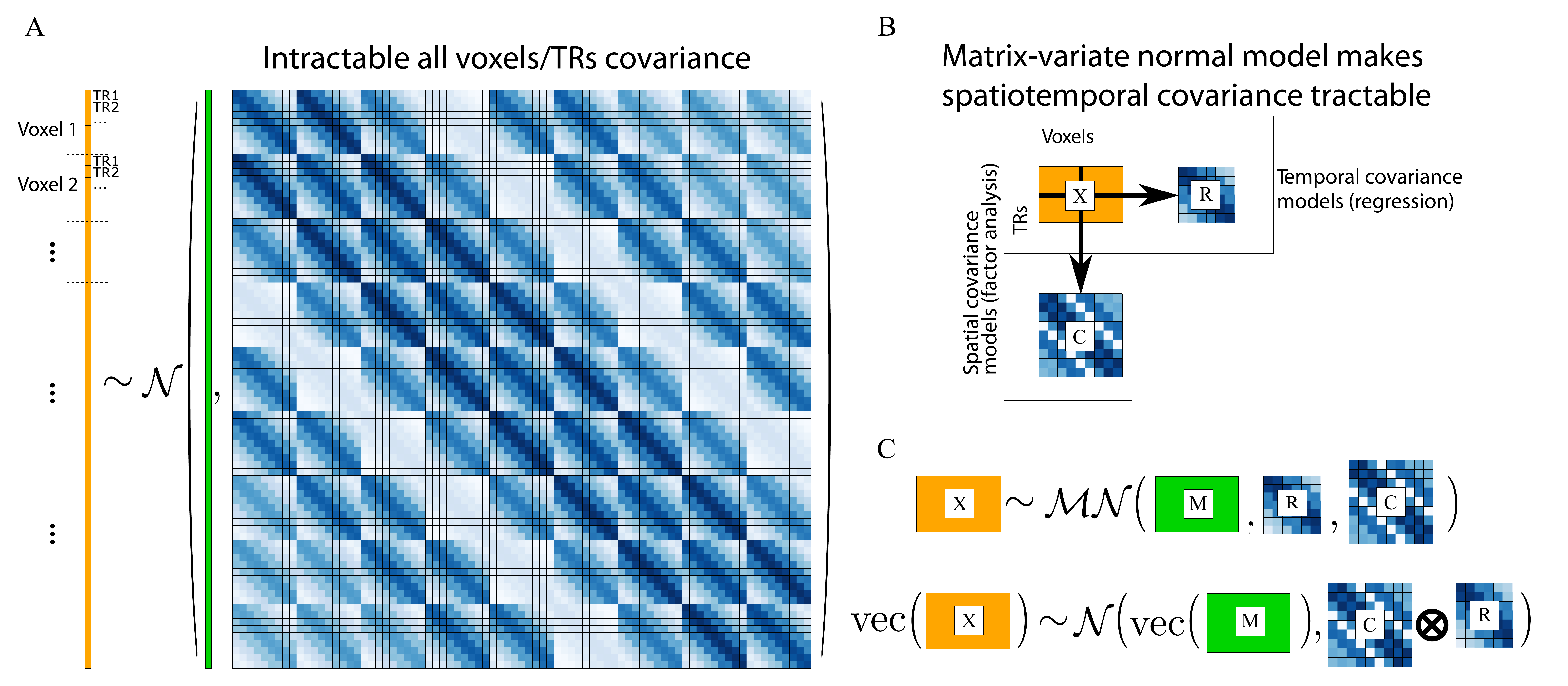}
    
    \caption{\textbf{Matrix normal models simultaneously model spatial and temporal residuals.} [A]: a schematic view of a vectorized data matrix, where each voxel's time series is vertically concatenated (in orange), and the covariance of every voxel at every timepoint with every other voxel at every other timepoint is modeled. Modeling all of these elements independently is intractable, and some structure needs to be imposed -- in this case, kronecker-separable structure. [B]: the un-vectorized data matrix (orange rectangle), and its spatial and temporal covariances on the right and bottom. [C]: A matrix-normal distribution with the mean M and row/column covariances R, C is equivalent to the large structure in [A], but can be much more tractable to estimate. }
    \label{fig:matnormal_schematic}
\end{figure}

As noted above, both the fMRI signal and residual are autocorrelated in both space and time; thus, modeling the residual structure in both dimensions is needed. This is not tractable in the general case, as it effectively means modeling the covariance between every voxel at every timepoint with every other voxel at every other timepoint. A simplifying assumption that permits modeling residuals in both space and time is that the spatial residuals of all time points have the same distribution, and the temporal residuals of all voxels likewise have the same distribution (for an illustration, see Fig.~\ref{fig:matnormal_schematic}). This \emph{separable} residuals assumption has been made in a GLS framework by~\citet{Katanoda2002} and factor-analytic framework by~\citet{Shvartsman2018}. A similar approach has been taken to modeling the entire dataset (rather than residuals only) in both neuroimaging~\citep[e.g.][]{Bijma2005,Ros2014} and elsewhere in the \emph{multitask learning} community~\citep[e.g][]{Bonilla2008,Skolidis2011,Stegle2011,Rakitsch2013,Greenewald2015}. 
Once separability is assumed, theoretically motivated structure could be placed on the individual spatial and temporal residual covariances, for example autoregressive in time (as in BRSA, above) and smooth in space (as in DRD, above). 

\vspace{0.1in}

\noindent\textsl{- Translating assumptions to a graphical model:  matrix-normal}

\begin{figure}[thb]
    \includegraphics[width=\linewidth]{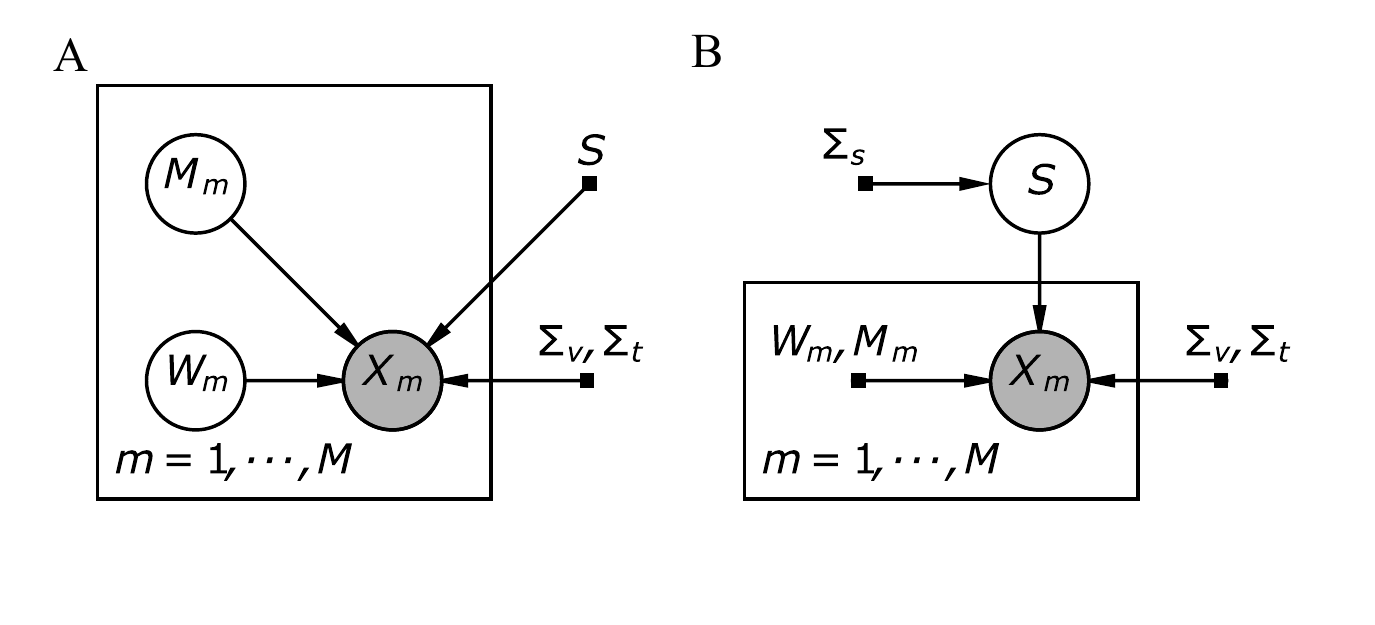}
    \caption{\textbf{Plate diagrams for matrix-normal shared response model.} In the matrix-normal notation one can see that there are two possible formulations for an SRM-type model: one which integrates over the shared timecourse (as SRM does), and one which integrates over the subject-specific weightings while removing the orthonormality assumption on $\mat{W_m}$ (this is termed `dual probabilistic SRM' or DP-SRM by analogy to dual probabilistic PCA, which makes the same extension to PCA \protect\citep{Lawrence2005}). In both cases, the brain image data $X_m\in\RR^{T\times V}$ is observed from subject $m$, $m = 1:M$. As in conventional SRM, each observation (now represented as the full data matrix) is a linear combination of subject-specific latent factors. In regular MN-SRM (A, left), the time-course $\mat{S}$ is treated as a latent variable that is integrated over and the mean $\mat{M}_m$ and weight vector $\mat{W}_m$ are treated as (hyper)parameters that need to be estimated. In DP-SRM (B, right), the weight vector and mean matrix are treated as latent variables and integrated over whereas the shared timecourse is treated as a (hyper)parameter to estimate. 
    Note how in contrast to \protect{Fig.~\ref{fig:SRM}}, there is no plate denoting independence between timepoints, since their covariance is now modeled. Shaded nodes: observations, unshaded nodes: latent variables, and black squares: parameters.}
    \label{fig:mn-srm}
\end{figure}
The informal claim of separability above is denoted by defining $\Sigma_{all}$ to be equal to the kronecker product of a spatial and temporal residual covariance, $\Sigma_{all}:=\Sigma_t \otimes \Sigma_v$. The kronecker product is a generalization of the vector outer product to matrices, and precisely performs the weighted tiling illustrated in Fig.~\ref{fig:matnormal_schematic}. Using this notation, we define 
the matrix-variate normal distribution, a distribution over matrices parameterized by a mean matrix and (separable) row and column covariances. We denote matrices drawn from this distribution as $\mat{X} \sim \mathcal{MN}_{m,n}(\mat{M},\mat{R},\mat{C})$, with mean $\mat{M}\in\mathbb{R}^{m\times n}$, row ccovariance $\mat{R}\in \mathbb{R}^{m \times m}$ and column covariance $\mat{C} \in \mathbb{R}^{n\times n}$. It has the following log-likelihood: 
\nopagebreak
\begin{align}
\log p(\mat{X}\mid \mat{M},\mat{R}, \mat{C}) =& -2\log mn - m \log|\mat{C}|
 \\
 &- n \log|\mat{R}| - \Tr\left[\mat{C}\inv(\mat{X}-\mat{M})\trp\mat{R}\inv(\mat{X}-\mat{M}M)\right]. \nonumber
\end{align}

The above notation is equivalent to denoting  $\vecop(\mat{X})\sim\mathcal{N}(\vecop(\mat{M}),\mat{C}\otimes\mat{R})$, where $\otimes$ is the kronecker product and $\vecop$ is the vectorization operator. If the column covariance $\mat{C}$ is the identity matrix (i.e.\ the columns are independent), the expression reduces to the log-likelihood of the multivariate normal distribution summed over the columns. We can use this notation to write, for example, 
a separable-residual model SRM model: 
\nopagebreak
\begin{align}
\mat{S} &\sim \mathcal{MN}(0, \mat{\Sigma_s}, \mat{I})\\
\mat{X}\trp_{m}\mid\mat{S} &\sim \mathcal{MN}(\mat{W_m}\mat{S}+\mat{M}_m, \mat{\Sigma_v}, \mat{\Sigma_t}), 
\end{align}
\nopagebreak
where $\mat{\Sigma_v}$ and $\mat{\Sigma_t}$ are spatial and temporal residual covariances and the remaining parameters are as defined above. In contrasting the diagram in Fig.~\ref{fig:mn-srm} one can see the disappearance of the plate iterating over timepoints, since now temporal residuals are modeled. In this view, we can also see a similar model in which the prior on $\mat{W}_m$ is modeled instead: 
\nopagebreak
\begin{align}
\mat{W_m} &\sim \mathcal{MN}(0, \mat{I}, \mat{\Sigma_w})\\
\mat{X}\trp_{m}\mid\mat{W_m} &\sim \mathcal{MN}(\mat{W_m}\mat{S}+\mat{M}_m, \mat{\Sigma_v}, \mat{\Sigma_t}). 
\end{align}

In this view, which Shvartsman et al.\ labeled dual probabilistic SRM (DP-SRM) by analogy to dual probabilistic PCA~\citep{Lawrence2005}, $\mat{W}_m$ can no longer be modeled as orthonormal but can now be integrated over with a gaussian prior, estimating substantially fewer parameters. 
Similar modeling of residual covariance can be performed on other factor models~\citep{Shvartsman2018}, including all of the generative models in this paper, or a generative variant of the structured sparsity (DRD) model, and others such as ISFC~\citep{Simony2016}. It is not obviously applicable to discriminative models, whose residuals are in the space of predictors and not the space of voxels. 

\vspace{0.1in}

\noindent\textsl{- Solving the model}

While simply estimating all parameters by gradient descent is theoretically possible, a more practical approach is to marginalize over nuisance parameters, and estimate only the parameters of interest. Marginalization in the multivariate normal setting with gaussian priors is well-known~\citep{Bishop2006}, but the separable covariance formulation introduces some new inference challenges: marginaliation yields a non-separable marginal likelihood, naive computation of which would require inverting a matrix of dimension $vt \times vt$ for $v$ voxels and $t$ time points, which is intractable for fMRI data. However,~\citet{Rakitsch2013} provided an efficient method for computing this likelihood by exploiting the compatibility between diagonalization and the kronecker product. If the spatial residual matrix itself needs to be separable (e.g.\ for efficiently modeling whole-brain spatial residuals by separating them in the x, y, and z dimensions),~\citet{Shvartsman2018} show that particular assumptions about prior covariances can likewise render the marginal separable (and thus tractable). Once the marginal likelihood can be computed efficiently, standard gradient-based techniques can be used for estimation. For even greater speed,~\citet{Shvartsman2018} derived an expectation-conditional-maximization algorithm for maximizing the marginal likelihood by coordinate ascent (though they only did so for matrix-normal SRM; matrix-normal RSA was estimated by gradient ascent).

\vspace{0.1in}

\noindent\textsl{Applications and benefits}

Similarly to the other models in the paper, we are not advocating any specific spatiotemporal covariance model here, nor its specific application to any specific method. Rather, we highlight the explicit modeling approach and its ability to incorporate a class of structure assumptions into other models, as long as they are linear Gaussian regression or factor models (which includes many models in the literature). 
That said, specific empirical benefits of introducing a separable residual covariance to other models have been realized.  In the case of the GLM for fMRI,~\citet{Katanoda2002} validated the separable-residuals model on synthetic data, as well as on a finger-tapping experiment. There, they demonstrated that the separable model recovers larger activations more closely focused around the expected motor regions. Additionally, the separable model provided a higher goodness of fit to experimental data than models that included temporal residual structure only, or no residual structure at all. In the case of factor models,~\citet{Shvartsman2018} show that the separable model can be substantially faster to estimate than a model that includes voxel-specific temporal residuals (as in the case of BRSA vs MN-RSA) and can achieve lower error while retaining BRSA's conservative behavior under the null. A separable variant of SRM achieves lower out-of-sample reconstruction error for new subjects than conventional SRM, though this reduced error does not seem to translate to improved feature extraction for brain decoding. The matrix-normal modeling toolkit (under review for inclusion in the BrainIAK analysis package~\citep{kumar2019brainiak}) makes it possible to prototype inclusion of separable covariances into other models.